\documentclass[final,5p,times,twocolumn]{elsarticle}

\usepackage{graphicx}

\usepackage{amsmath, amssymb}
\usepackage{epstopdf}
\usepackage{lmodern}
\usepackage[latin9]{inputenc}
\usepackage[T1]{fontenc}
\usepackage{color}
\usepackage{float}
\usepackage{tabularx,ragged2e,booktabs,caption}
\usepackage{breqn}
\usepackage{pgfplots}
\pgfplotsset{compat=1.5}
\usepackage{subcaption}

\biboptions{authoryear}

\journal{International Journal of Solids and Structures}

\begin{document}

\begin{frontmatter}


\title{Surface Tension Effects on Surface Instabilities of Dielectric Elastomers}


\author[ss]{Saman Seifi}
\ead[ss]{samansei@bu.edu}
\address[ss]{Department of Mechanical Engineering, Boston University, Boston, MA 02215}
\author[qm]{Qiming Wang}
\address[qm]{Department of
Civil and Environmental Engineering,
University of Southern California, Los Angeles, CA 90089}
\ead[qm]{qimingw@usc.edu}
\author[hsp]{Harold S. Park}
\ead[hsp]{parkhs@bu.edu}
\address[hsp]{Department of Mechanical Engineering, Boston University, Boston, MA 02215}


\begin{abstract}
Dielectric elastomers have recently been proposed for various biologically-relevant applications, in which they may operate in fluidic environments where surface tension effects may have a significant effect on their stability and reliability.  Here, we present a theoretical analysis coupled with computational modeling for a generalized electromechanical analysis of surface stability in dielectric elastomers accounting for surface tension effects.  For mechanically deformed elastomers, significant increases in critical strain and instability wavelength are observed for small elastocapillary numbers.  When the elastomers are deformed electrostatically, both surface tension and the amount of pre-compression are found to substantially increase the critical electric field while decreasing the instability wavelength.  
\end{abstract}


\end{frontmatter}


\section{Introduction}

Because soft materials like gels and elastomers often operate under states of compression, there have been significant efforts to examine their mechanical stability under such loading~\citep{Biot1963,gent1999surface,biot1965mechanics,tanaka1987mechanical,gent2005elastic,hong2009formation}.  Recently, efforts have focused on examining the stability of surfaces due to the formation of instabilities such as wrinkles and creases under compression~\citep{cao2012wrinkling,li2012mechanics,jin2014creases,cai2012creasing,hong2013crease,weiss2013creases}.  The studies of wrinkling instabilities have typically been analytical in nature, based on linear perturbation analysis, which enables predictions of critical strains and instability wavelengths in these soft structures.  These studies have also been insightfully combined with experiments~\citep{cai2010osmotic,cai2012creasing,Jin2015a}, and also numerical (finite element) analyses~\citep{cao2012wrinkling,jin2015smoothening,Jin2015a,wangJAM2014,wangMRS2016}.  

More recently, linear perturbation analyses have been applied to an interesting class of soft, active materials - dielectric elastomers (DEs).  These materials have drawn significant interest due to the large deformations they can undergo resulting from applied electrostatic loading~\citep{carpiSCIENCE2010,brochuMRC2010,biddissMEP2008,mirMT2007}.  Furthermore, DEs often form surface instabilities like creasing, wrinkling and cratering, which have been studied via experiment, theory and computation~\citep{suoAMSS2010,planteIJSS2006,zhouIJSS2008,wangAM2012,shivapoojaAM2013,wangPRL2011a,wang2013creasing,parkIJSS2012,parkCMAME2013}, and also using analytic techniques based on stability analyses~\citep{zhaoAPL2007,zhao2009electromechanical} and the nonlinear field theory of~\citet{suoJMPS2008}.

While DEs have drawn significant interest in recent years, many promising technological innovations, such as generating electromechanical motion during magnetic resonance imaging~\citep{Carpi2008mri}, operating soft, underwater robots~\citep{rusNATURE2015,kimTB2013}, manipulating microfluidic flow~\citep{Holmes2013}, focusing a tunable lens~\citep{Carpi2011}, the preparation of bioinspired surfaces~\citep{shivapoojaAM2013}, underwater grippers~\citep{Laschi2012} and soft body locomotion~\citep{Marchese2014}, involve operation of the DE in fluidic environments, where elastocapillary effects due to surface tension can become prominent~\citep{romanJPCM2010,liuAMS2012}.  While prior works have examined either the effects of surface tension alone on surface instabilities in elastomers~\citep{chenPRL2012} and hydrogels~\citep{kangSM2010}, or the coupled effects of surface tension and electric fields on surface instabilities in constrained DE films~\citep{wang2013creasing,seifiIJSS2016,wangJMPS2016}, a general stability analysis of compressed DE films subject to both surface tension and electric fields has not been performed.  

The objective of this work is to present a theoretical analysis coupled with computational modeling for surface tension effects on surface instabilities in DEs, with particular emphasis on surface instability wavelengths, and critical strains or electric fields needed to induce the surface instabilities.  We consider first the case of surface tension effects on a mechanically compressed DE film, then move on to consider the general case of electrostatic loading on a mechanically compressed DE film.  

\section{Governing Equations and Incremental Forms}
\label{AppexA}

We first present, for a DE film subject to surface tension and under compressive loading, the governing equations and their incremental forms.  We assume that the surface of the film is subject to Young-Laplace boundary conditions in order to account for elastocapillary effects on the surface~\citep{saksonoCM2006}.  We initially neglect the effects of applied electrostatic loading (i.e. voltages or electric fields), which will be accounted for in the subsequent section. The perturbation analysis we utilize is based on those performed by \citet{cai2010osmotic,kangSM2010,jin2014creases,Jin2015a}

\subsection{Dielectric Elastomer Subject to Surface Tension}

\begin{figure}
	\centering
	\includegraphics[scale=0.9]{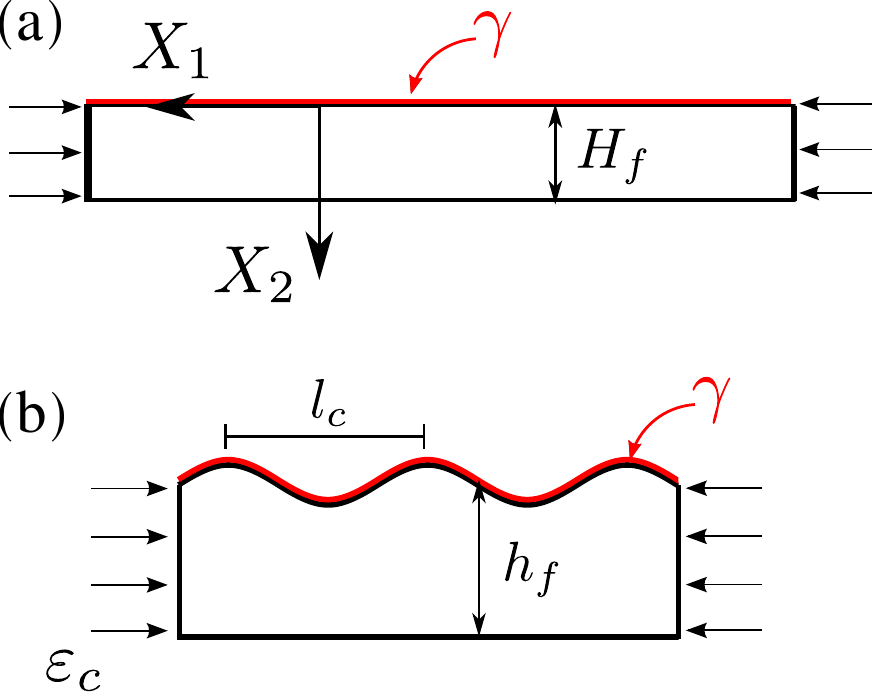}
	\caption{A perfectly flat film subject to compression and surface tension $\gamma$. (a) Undeformed configuration; (b) the formation of wrinkles with wavelength $l_c$. }
	\label{fig:mechfig}
\end{figure}

We first consider a DE film under compression, that is also subject to surface tension effects on its top surface, as illustrated in Figure (\ref{fig:mechfig}). The deformation of the body is described by the field $\mathbf{x}(\mathbf{X})$. The deformation gradient is:
\begin{equation}
	F_{iK}=\frac{\partial x_{i}(\mathbf{X})}{\partial{X_{K}}}
	\label{eq:defgrad1}
\end{equation}
The DE is modeled as an incompressible neo-Hookean material with the free energy function:
\begin{equation}
	W(\mathbf{F})=\frac{\mu}{2}F_{iK}F_{iK}-\pi(\det{\mathbf{F}}-1)
	\label{eq:neoHookean}
\end{equation}
The nominal stress $s_{iK}$ can be derived as:
\begin{equation}
	s_{iK}=\frac{\partial W(\mathbf{F})}{\partial F_{iK}}=\frac{\mu}{J}F_{iK}-\pi H_{iK}
	\label{eq:mechstress1}
\end{equation}
where $\mathbf{H}=\mathbf{F}^{-T}$, $\mu$ is the shear modulus, and $\pi=\pi(\mathbf{X})$ is the Lagrange multiplier to enforce the constraint of incompressibility: $J=\det{\mathbf{F}}=1$. 
The nominal stress satisfies the equilibrium equation:
\begin{equation}
	\frac{\partial s_{iK}}{\partial X_K}=0
	\label{eq:equil_equ}
\end{equation}

On the surface of the body the film is subjected to the Young-Laplace boundary condition~\citep{saksonoCM2006} with surface tension $\gamma$:
\begin{equation}
	s_{iK}N_{K}dA=2\gamma\kappa n_{i} da
	\label{eq:YLBC}
\end{equation}
where $\mathbf{N}$ is the unit normal vector to the surface of the body in the undeformed state, $\mathbf{n}$ is the unit normal vector in the current state, and $\kappa$ is the mean curvature of the surface. 

We now add a perturbation into the state of finite deformation with position $x^{0}_{i}(\mathbf{X})$, deformation gradient $F^{0}_{iK}(\mathbf{X})$ and nominal stress $s^{0}_{iK}(\mathbf{X})$.  If the deformation in the $z$-direction is constrained, only 2D deformation is considered.  The deformation gradient of this state is
\begin{equation}
\mathbf{F}^{0}=
	\begin{bmatrix}
		\lambda & 0  \\
		0 & 1/\lambda \\
	\end{bmatrix}
\end{equation}
Therefore the corresponding position is $\mathbf{x}^0 = \mathbf{F}^{0}\mathbf{X}$. Now by adding a small displacement perturbation of $\dot{\mathbf{x}}$, we write the perturbed state as $\mathbf{x} = \mathbf{x}^{0}(\mathbf{X})+ \dot{\mathbf{x}}(\mathbf{X})$. Both state $\mathbf{x}^{0}(\mathbf{X})$ and the perturbed state $\mathbf{x}^{0}(\mathbf{X} )+\dot{\mathbf{x}}(\mathbf{X})$ satisfy the same governing equations in (\ref{eq:defgrad1}) and (\ref{eq:equil_equ}). The incremental form of the deformation gradient reduces to
\begin{equation}
	\dot{F}_{iK}=\frac{\partial \dot{x}_{i}(\mathbf{X})}{\partial X_K}
	\label{eq:defgraddot}	
\end{equation}
and the equilibrium equation to
\begin{equation}
	\frac{\partial \dot{s}_{iK}}{\partial X_K}=0
	\label{eq:dot_equilbrium}
\end{equation}
where the incremental nominal stress can be obtained using Taylor's expansion at $\mathbf{F}^{0}$ of the stress in (\ref{eq:mechstress1}):
\begin{equation}
	\dot{s}_{iK}=\mu \dot{F}_{iK}-\dot{\pi} H^{0}_{iK}+\pi^{0} H^{0}_{iL} H^{0}_{jK}\dot{F}_{jL}
	\label{eq:Sigma_ij}
\end{equation}
Inserting (\ref{eq:Sigma_ij}) into (\ref{eq:dot_equilbrium}) we can obtain the following equation:
\begin{equation}
	\left(\mu \delta_{ij}\delta_{KL}+\pi^{0}H^{0}_{iL}H^{0}_{jK}\right)\frac{\partial^{2} \dot{x}_j(\mathbf{X})}{\partial X_K \partial X_L}-H^{0}_{iK}\frac{\partial \dot{\pi}}{\partial X_K}=0
	\label{eq:dotequil1}
\end{equation}
The perturbation of the incompressibility condition becomes
\begin{equation}
	H^{0}_{iK}\dot{F}_{iK}=0
	\label{eq:icompressdot}
\end{equation}
where the perturbation of the boundary-condition is:
\begin{equation}
	\dot{s}_{iK}N_{K}=2\gamma\dot{\kappa}H^{0}_{iK}N_{K}+2\gamma\kappa \dot{H}^{0}_{iK}N_{K}
	\label{eq:YPBCdot}
\end{equation}
Finally, the boundary condition on the top the film (\ref{eq:YPBCdot}) reduces to
\begin{equation}
	\dot{s}_{iK}N_K=2\gamma\dot{\kappa}H^{0}_{iK}N_K 
	\label{eq:BC_gamma}
\end{equation}
where the term involving $\kappa$ in (\ref{eq:YPBCdot}) drops out as $\kappa=0$.  At the bottom of the film, we have the boundary conditions:
\begin{equation}
	\dot{s}_{12}=0 \quad \text{and} \quad \dot{x}_2=0
	\label{eq:BC_bottom1}
\end{equation}

\subsection{Dielectric Elastomer with Surface Tension and Electric Field Effects}

\begin{figure}
	\centering
	\includegraphics[scale=0.9]{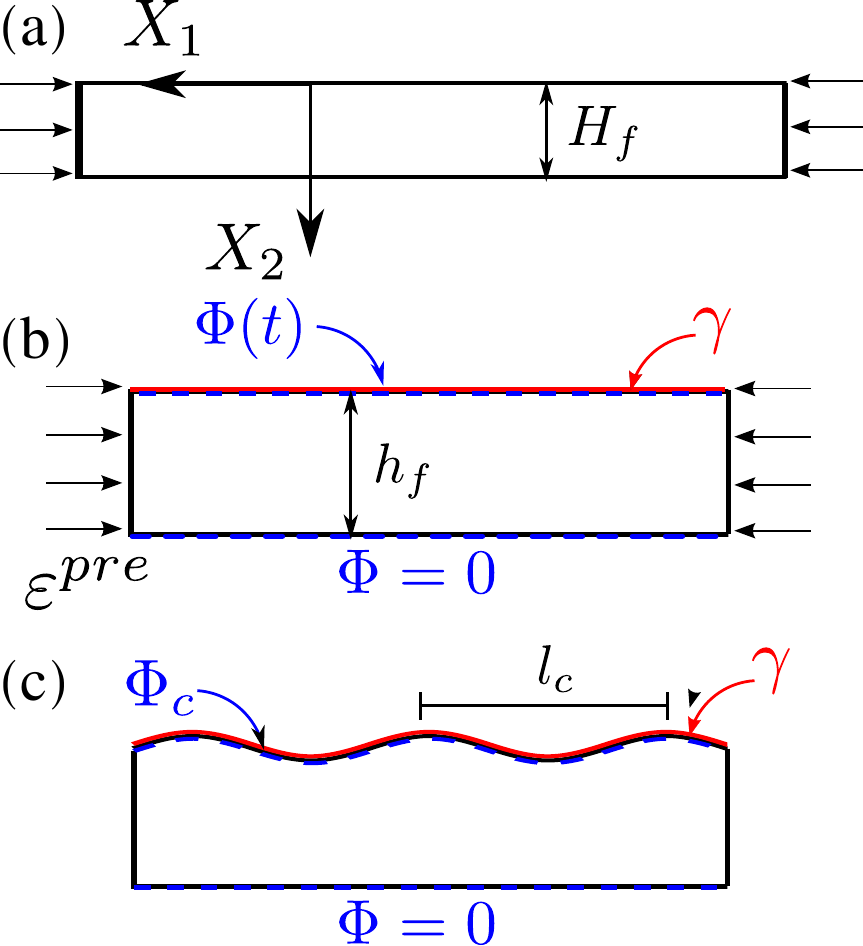}
	\caption{A perfectly flat film subject to compression, electric field and surface tension $\gamma$. (a) Undeformed configuration; (b) Compressed to a given strain $\varepsilon^{pre}$, after which voltage $\Phi(t)$ is applied to the top surface while $\Phi=0$ on the bottom surface; (c) The formation of wrinkles with critical wavelength $l_c$ at critical electric potential $\Phi_c$. }
	\label{fig:elecmechfig}
\end{figure}

We now consider the general problem of interest, that of a DE film that is subject to compressive loading, while accounting for both surface tension and electrostatic loading via electric fields.  This is illustrated in Figure (\ref{fig:elecmechfig}). In this case, the reference state is a pre-compressed plane strain film, subject to the deformation gradient 
\begin{equation}
\mathbf{F}^{pre}=
	\begin{bmatrix}
		\lambda^{pre} & 0  \\
		0 & 1/\lambda^{pre} \\
	\end{bmatrix}
\end{equation}
For this film, the net deformation gradient between the current state and the undeformed state is $F_{iK} F^{pre}_{Km}$, and the free energy of equation in (\ref{eq:neoHookean}) becomes
\begin{equation}
	W(\mathbf{F})=\frac{\mu}{2}F_{iK}F^{pre}_{Km}F_{iL}F^{pre}_{Lm}-\pi(\det{\mathbf{F}}-1)
	\label{eq:neHookean_pre}
\end{equation}
and therefore the nominal stress of (\ref{eq:mechstress1}) becomes
\begin{equation}
	s_{iK}=\frac{\mu}{J}F^{pre}_{Km}F^{pre}_{Lm}F_{iL}-\pi H_{iK}.
	\label{eq:mechstress_pre}
\end{equation}
At this state (Figure \ref{fig:elecmechfig}-b), an electric potential $\Phi$ is applied on top of the film along with the surface tension $\gamma$. When the elastomer is at the flat state, the electric field in the elastomer film is homogeneous, with the electric field vector as 
\begin{equation}
	\mathbf{E}=\begin{bmatrix}
	0 \\
	E_2 \\
\end{bmatrix}	
	\label{eq:elecfield} 
\end{equation}
where $E_2=\Phi/h_f$ is the applied electric field where $h_f=H_f/\lambda^{pre}$ is the height of the pre-compressed film. 

The electric field $\mathbf{E}$ generates an extra stress $\mathbf{s}^{e}$ inside the film. The nominal electric stress can be calculated as $\mathbf{s}^{e}=J {\sigma}^{e}\mathbf{H}^{0}$ where $\sigma^{e}=\epsilon\mathbf{E}\otimes \mathbf{E}-\frac{\epsilon}{2}|\mathbf{E}|^{2} \mathbf{I}$. Also here $\mathbf{H}^{0}=(\mathbf{F}^0)^{-T}$ where $\mathbf{F}^{0}$ is the deformation gradient that describes the post-compressed motion. Therefore we can obtain the nominal electric stress as
\begin{equation}
	\mathbf{s}^{e}=J \begin{bmatrix}
	-\frac{1}{2}\epsilon E^{2}_{2} & 0 \\
	0 & \frac{1}{2}\epsilon E^{2}_{2}
	\end{bmatrix}\mathbf{H}^{0}.
	\label{eq:se1}
\end{equation}
The equilibrium equation becomes
\begin{equation}
	\frac{\partial (s_{iK}+s^{e}_{iK})}{\partial X_{K}}=0
\end{equation}
With relation (\ref{eq:se1}), and the fact that the film remains flat before the onset of surface wrinkling, we have the approximation $\mathbf{F}^{0}\approx\mathbf{I}$ and therefore $(\mathbf{F}^0)^{-T}= \mathbf{H}^{0}\approx\mathbf{I}$, we can obtain $\frac{\partial s^{e}_{iK}}{\partial X_{K}}\approx0$. Thus the the stress equilibrium reduces to:
\begin{equation}
	\frac{\partial s_{iK}}{\partial X_{K}}	=0
\end{equation}
Now similar to the previous case, we perturb the post-compressed state of deformation, of displacement $x^{0}_{i}$ and deformation gradient $\mathbf{F}^{0}$.  The incremental state of stress becomes:
\begin{equation}
	\dot{s}_{iK}=\mu F^{pre}_{Km}F^{pre}_{Lm}\dot{F}_{jL}+\pi^{0} H^{0}_{iL}H^{0}_{jK}\dot{F}_{jL}-H^{0}_{iK}\dot{\pi}	
\end{equation}
and the equilibrium equation becomes:
\begin{equation}
	\frac{\partial \dot{s}_{iK}}{\partial X_K}=0
	\label{eq:equil_2}
\end{equation}
where it can be rewritten as:
\begin{equation}
	\left(\mu F^{pre}_{Km}F^{pre}_{Lm}\delta_{ij}+\pi^{0}H^{0}_{iL}H^{0}_{jK}\right)\frac{\partial^2 x_{j}}{\partial X_K \partial X_L}-H^{0}_{iL}\frac{\partial\dot{\pi}}{\partial X_K}=0
	\label{eq:equil_2_pre}
\end{equation}
After rearrangement of electric stress the boundary condition on top of the film is:
\begin{equation}
	\dot{s}_{iK}N_{K}=2\gamma\dot{\kappa}n_i-\dot{s}^{e}_{iK}N_K
	\label{eq:BC_elec_gamma}
\end{equation}
where according to (\ref{eq:se1}) we have $\dot{s}^{e}_{iK}=\dot{\sigma}^{e}_{ij}$ as a consequence of the fact that $\mathbf{H}^{0}\approx\mathbf{I}$. By assuming the fact that the DE film undergoes a sinusoidal undulation with small amplitude $\delta$ at the onset of wrinkling, the electric field then is $(E_2)_{\delta}=\Phi/(h_f+\delta)$, and therefore the linear approximation of the electrical stress using Taylor's series becomes: 
\begin{equation}
	\dot{{\sigma}}^{e}=\begin{bmatrix}
	\epsilon E^{2}_{2}\left(\frac{\delta}{h_f} \right) & 0 \\
	0 & -\epsilon E^{2}_{2}\left(\frac{\delta}{h_f} \right)
	\end{bmatrix}
\end{equation}
And finally the boundary conditions at the bottom of the film is fixed:
\begin{equation}
	\dot{x}_1(X_1, H_f)=\dot{x}_2(X_1,H_f)=0
	\label{eq:BC_bottom2}
\end{equation}

\section{Linear Perturbation Analysis for Wrinkles}
\label{AppexB}
\subsection{Compression with elastocapillary effect}

\begin{figure}
	\centering
	\includegraphics[scale=0.9]{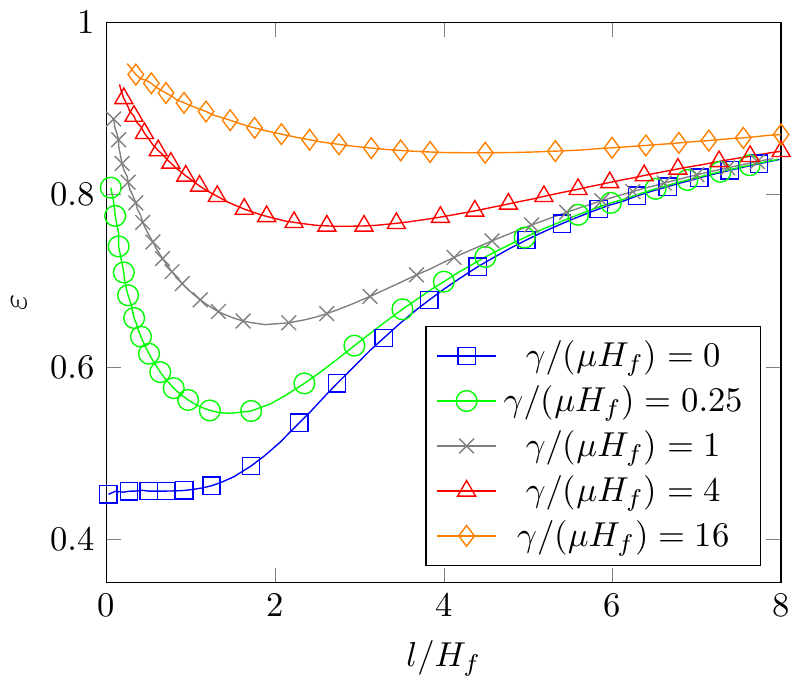}
	\caption{Strain vs. wavelength for a range of elastocapillary numbers}
	\label{fig:mechfig_plot}
\end{figure}

\begin{figure}
	\centering
	\begin{subfigure}[b]{0.425\textwidth}
		\includegraphics[scale=0.9]{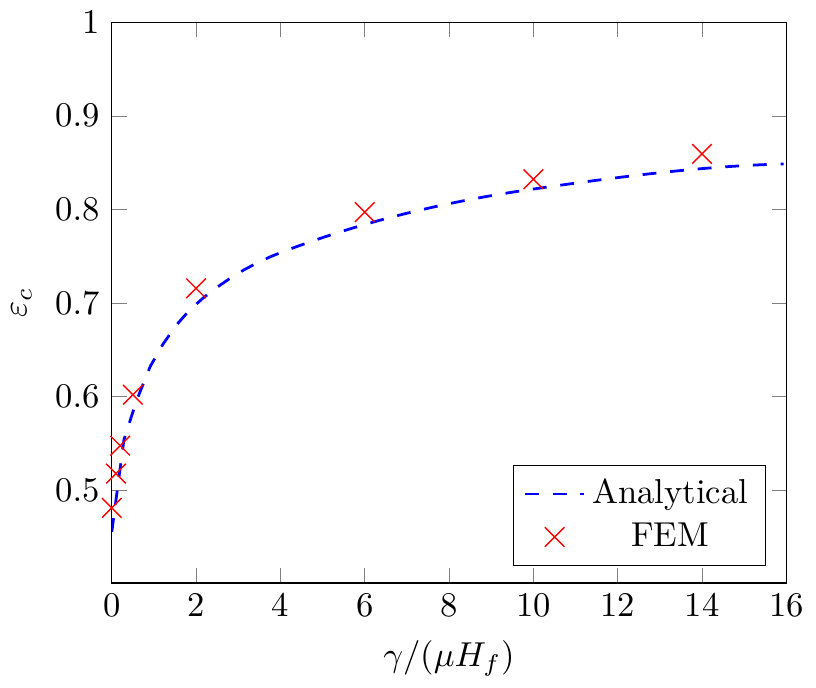}
		\caption{}
	\end{subfigure}
	\begin{subfigure}[b]{0.42\textwidth}
		\includegraphics[scale=0.9]{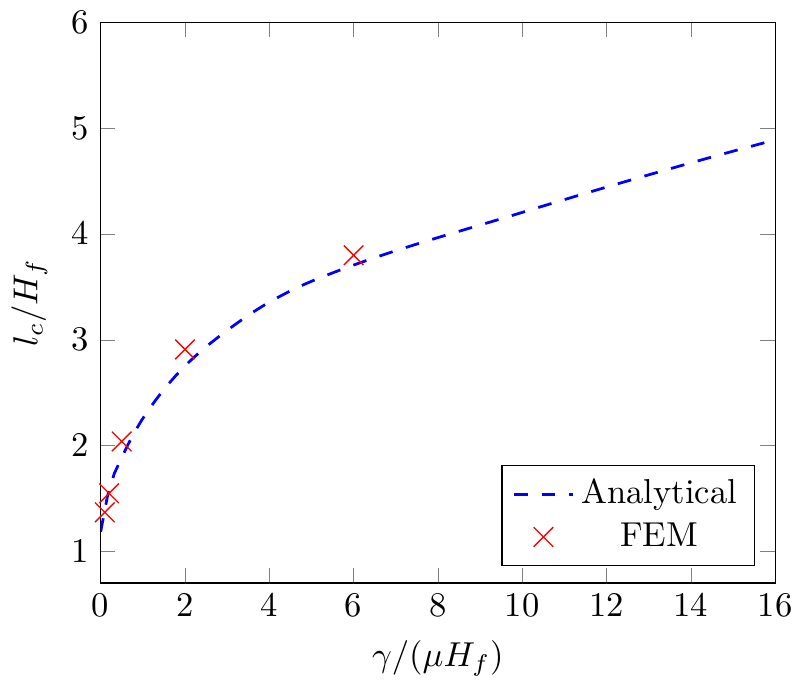}
		\caption{}
	\end{subfigure}
	\caption{(a):  Critical strain $\varepsilon_c$ vs. elasto-capillary number $\gamma/(\mu H_f)$; (b): Critical wavelength $l_c/H_f$ vs. elasto-capillary number $\gamma/(\mu H_f)$}
	\label{fig:mechfig_plot_fem}
\end{figure}

To determine the onset of wrinkling, we require the existence of non-trivial solutions to the eigenvalue problem corresponding to the incremental boundary value problems derived in the previous section. 
Separated solution exists in the incremental boundary value problem with the perturbation in the following form:
\begin{equation}
	\begin{aligned}
		\dot{x}_{1}(X_1,X_2)=f_{1}(X_2)\sin( K X_1) \\
		\dot{x}_{2}(X_1,X_2)=f_{2}(X_2)\cos(K X_1) \\
		\dot{\pi}(X_1,X_2)=f_{3}(X_2)\cos(K X_1)
	\end{aligned}
	\label{eq:per_sol}
\end{equation}
Substituting these equations into (\ref{eq:dotequil1}) we obtain
\begin{equation}
	f''''_{2}-	K^2\left(\lambda^{-4}+1\right)f''_{2}+{K^4}{{\lambda}^{-4}}f=0
		\label{eq:ODE_2}
\end{equation}
The derivatives of $f_2$ are with respect to $X_2$ and $K=2\pi/L$ is the wave number. The wavelength in the reference state $L$ relates to wavelength in current $l$ state by $L=l/\lambda$. The ODE in (\ref{eq:ODE_2}), accompanied with the boundary conditions in (\ref{eq:BC_gamma}) and (\ref{eq:BC_bottom1}), leads to an eigenvalue problem, of which the non-trivial solutions correspond to the wrinkling state. The corresponding algebraic equation of the following form then can be obtained

\begin{equation}
	\varmathbb{A}\begin{bmatrix}
	C_1 \\
	C_2 \\
	C_3 \\
	C_4 
	\end{bmatrix}=0
	\label{eq:eigen_1}
\end{equation}
The existence of a non-trivial solution to the perturbed boundary value problem requires
\begin{equation}
	\det \varmathbb{A}=0
	\label{eq:eig_1}
\end{equation}
where matrix $\det\varmathbb{A}$ can be written as a function of $\det\varmathbb{A}=f(\lambda,K H_f,\gamma/(\mu H_f))$. The explicit expression of this matrix is given as follows:
\begin{equation}
\begin{gathered}
	A_{11}=-\frac{\gamma  K H_f}{\lambda \mu }- H_f K\left(\lambda^2  +\frac{  1}{\lambda ^2}\right) \\
	A_{12}=-\frac{\gamma  K H_f}{\lambda \mu }+ H_f K\left(\lambda^2  +\frac{  1}{\lambda ^2}\right) \\
	A_{13}=-\frac{\gamma H_f K^2}{\mu \lambda }-2   K H_f \\
	A_{14}=-\frac{\gamma H_f K^2}{\mu \lambda }+2   K H_f  
	\end{gathered}
\end{equation}
\begin{equation}
	\begin{gathered}
	A_{21}=A_{22}=\frac{2 K H_f}{\lambda ^2} \\
	A_{23}=A_{24}= \lambda ^2 K H_f+\frac{K H_f}{\lambda ^2}
	\end{gathered}
\end{equation}	
\begin{equation}
	\begin{gathered}
	A_{31}=\frac{2 K H_f \,e^{-K H_f {\lambda ^{-2}}}}{\lambda ^2} \\
	A_{32}=\frac{2 K H_f \,e^{K H_f {\lambda ^{-2}}}}{\lambda ^2} \\
	A_{33}=\lambda ^2 K H_f e^{-K H_f}+\frac{ K H_f e^{-K H_f}}{\lambda ^2} \\
	A_{34}=\lambda ^2 K H_f e^{K H_f}+\frac{ K H_f e^{K H_f}}{\lambda ^2}
	\end{gathered}
\end{equation}	
\begin{equation}
	\begin{gathered}
	A_{41}=e^{-{K H_f}{\lambda ^{-2}}}\ ,\ A_{42}=e^{{K H_f}{\lambda ^{-2}}}\\
	 A_{43}=e^{-K H_f}\ , \ A_{44}=e^{K H_f}
	\end{gathered}
\end{equation}
By solving the eigenvalue problem in (\ref{eq:eig_1}), we obtain the relation between stretch $\lambda$, and therefore the compressive strain $\varepsilon=1-\lambda$ with the wavelength $l$ (Figure \ref{fig:mechfig_plot}). For each curve, the minimal critical strain $\varepsilon_c=1-\lambda_c$ reaches a minimum for wrinkles of certain wavelength $l_c$. The relation of these critical values with elastocapillary number $\gamma/(\mu H_f)$ is plotted in Figure \ref{fig:mechfig_plot_fem}.

When surface tension is neglected ($\gamma=0$), we recover  Biot's value of wrinkling at $\varepsilon_{w}\approx0.46$ as shown in Figure \ref{fig:mechfig_plot}~\citep{Biot1963}.  When surface tension is present ($\gamma>0$), the strain $\varepsilon$ first decreases and then increases as the normalized wavelength $l/H_f$ increases (Figure \ref{fig:mechfig_plot}).  The key effect of surface tension is that it inhibits bifurcation, as both the critical strain $\varepsilon_c$ and corresponding normalized wavelength $l_c/H_f$ increase with increasing $\gamma/(\mu H_f)$ (Figure \ref{fig:mechfig_plot_fem}).  These results are consistent with previous studies on surface tension effects on surface instabilities~\citep{chenPRL2012}, and are expected since a smaller wavelength $l/H_f$ implies a larger energy penalty in terms of surface energy \citep{Shenoy2001}.  

In addition, we find that for large elastocapillary numbers, the critical strain approaches a limiting value of $\varepsilon_{c}\approx0.85$.  The largest change in critical strain occurs for small elastocapillary numbers, i.e. $0\leqslant\gamma/(\mu H_f)\leqslant 2$, where the change in $\varepsilon_c$ is more than $34\%$ from $\varepsilon_c=0.46$ for $\gamma/(\mu H_f)=0$ to $\varepsilon_c\approx0.7$ for $\gamma/(\mu H_f)=2$, while for $\gamma/(\mu H_f)>2$ the increase in $\varepsilon_c$ is less than $18\%$  (Figure \ref{fig:mechfig_plot_fem}a). 

The solutions to the critical wavelength are different from the critical strain in that there is no limit to the wavelength.  As with the critical strain in Figure \ref{fig:mechfig_plot_fem}a, a rapid increase in normalized wavelength is observed for small elastocapillary numbers.  For $\gamma/(\mu H_f)>4$, where the change in critical strain is not significant, the normalized wavelength continues increasing in a linear fashion. 

We also verified the theoretical model by performing dynamic, nonlinear finite element (FE) calculations using the methodology for electro-elastocapillary phenomena in DEs previously developed by~\citet{seifiIJSS2016}, while neglecting the electrostatic effects, thus considering a purely mechanical problem.  All numerical simulations using open source simulation code~\citet{tahoe} using standard 4-node, bilinear quadrilateral finite elements within a two-dimensional, plane strain approximation. Simulations were performed on a elastomer film with length $L_f=40$ and height of $H_f=4$ with a mesh size of $d=1/16$.

\begin{figure}
	\centering
	\includegraphics[scale=0.3]{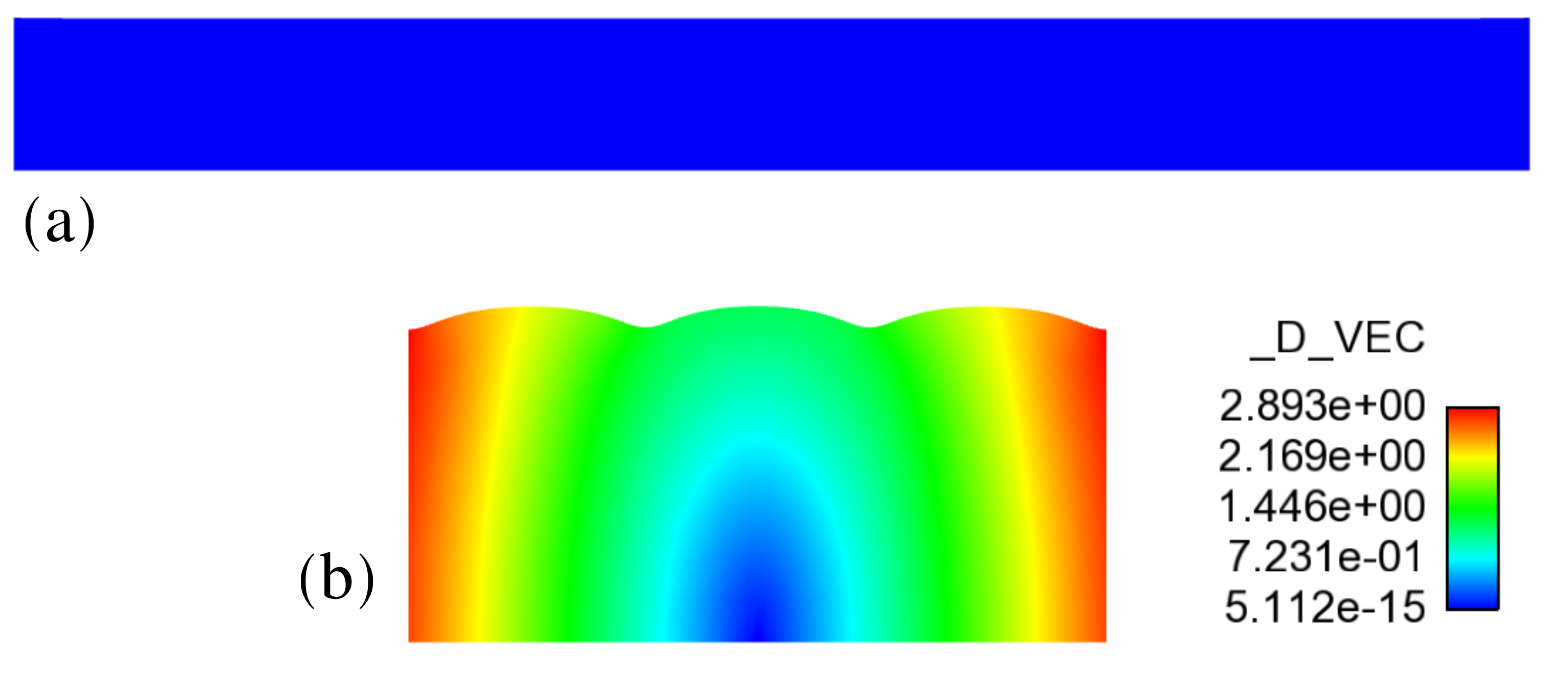}
\caption{Results for FE simulation of a mechanical compression-induced instability of an $40\times4$ elastomer film. (a) Undeformed film (b) Onset of wrinkling instability on the top surface at the critical strain $\varepsilon_c=0.547$ and elastocapillary number $\gamma/(\mu H_f)=0.2$. _D_VEC is the displacement magnitude.}
	\label{fig:fem_s12}
\end{figure}

In the FE simulations, we measured the critical strain $\varepsilon_c$ as soon as the wrinkling pattern on surface appears, see for example (Figure \ref{fig:fem_s12}).  For the comparisons to the critical strain in Figure \ref{fig:mechfig_plot_fem}(a), the FE results match the theoretical model very closely.  For the normalized wavelength $l_{c}/H_{f}$ in Figure \ref{fig:mechfig_plot_fem}(b), the FE results match also closely match the theoretical model.  However, we were not able to obtain the wavelength of the wrinkles for $\gamma/(\mu H_f)>2$ where the film is more than $\%70$ compressed due to the computational expense needed in modeling very long films.

\subsection{Compressed film subject to surface tension and electric field}  

To account for electromechanical coupling on the compression-induced instability, we substitute equations in (\ref{eq:per_sol}) into  (\ref{eq:equil_2_pre}) giving the following differential equation:
\begin{equation}
	f''''_2-K^2\left((\lambda^{pre})^{4}+1 \right)f''_2+K^4 (\lambda^{pre})^{4}f_2=0.
	\label{eq:ODE_3}
\end{equation}
This equation along with the boundary conditions in (\ref{eq:BC_elec_gamma}) and (\ref{eq:BC_bottom2}) gives the second algebraic equation:
\begin{equation}
	\varmathbb{B}\begin{bmatrix}
	C_1 \\
	C_2 \\
	C_3 \\
	C_4 
	\end{bmatrix}=0
	\label{eq:eigen_2}
\end{equation}
The existence of a non-trivial solution requires:
\begin{equation}
	\det{\varmathbb{B}}=0
	\label{eq:eig_2}
\end{equation}
where the matrix $\det\varmathbb{B}$ can be written as function of dimensionless parameters $\det\varmathbb{B}=g(K H_f, \tilde{E}\sqrt{\epsilon/\mu}, \gamma/(\mu H_f) )$ and the stretch $\lambda^{pre}$ is a prescribed constant and where $\tilde{E}=\Phi/H_f$ is the nominal electric field. The explicit expression of the matrix $\varmathbb{B}$ is:
\begin{equation}
	\begin{gathered}
		B_{11}=-\frac{(\lambda^{pre}) ^2 \Phi^2 \epsilon }{\mu H_{f}^{2}}-\frac{\gamma H_f K^2}{\mu (\lambda^{pre}) }-2  K H_f \\
		B_{12}=-\frac{(\lambda^{pre}) ^2 \Phi^2 \epsilon }{\mu H_{f}^{2}}-\frac{\gamma H_f K^2}{\mu (\lambda^{pre}) }+2  K H_f \\
		B_{13}=-\frac{(\lambda^{pre}) ^2 \Phi^2 \epsilon }{\mu H_f^2}-\frac{\gamma H_f K^2}{\mu (\lambda^{pre}) }\\ - K H_f\left((\lambda^{pre}) ^2 +\frac{1}{(\lambda^{pre}) ^2}\right)\\
		B_{14}=-\frac{(\lambda^{pre}) ^2 \Phi^2 \epsilon }{\mu H_f^2}-\frac{\gamma H_f K^2}{\mu (\lambda^{pre}) }\\ + K H_f\left((\lambda^{pre}) ^2 +\frac{1}{(\lambda^{pre}) ^2}\right)
	\end{gathered}
\end{equation}
\begin{equation}
	\begin{gathered}
		B_{21}=B_{22} = -(\lambda^{pre}) ^2 K H_f-\frac{K H_f}{(\lambda^{pre}) ^2} \\
		B_{23}=B_{34} = -\frac{2 K H_f}{(\lambda^{pre}) ^2}
	\end{gathered}
\end{equation}
\begin{equation}
	\begin{gathered}
		B_{31} = (\lambda^{pre}) ^2 K H_f \left(-e^{-H_f  K (\lambda^{pre})}\right) \\
		B_{32} = (\lambda^{pre}) ^2 K H_f \left(e^{H_f  K (\lambda^{pre})}\right) \\
		B_{32} = K H_f \left(-e^{-\frac{K H_f}{(\lambda^{pre}) }}\right) \ , \
		B_{34} = K H_f \left(e^{\frac{K H_f}{(\lambda^{pre}) }}\right) 
	\end{gathered}
\end{equation}
\begin{equation}
	\begin{gathered}
		B_{41} = e^{-K H_f (\lambda^{pre})} \ , \
		B_{42} = e^{K H_f (\lambda^{pre})} \\
		B_{43} = e^{-\frac{K H_f}{(\lambda^{pre}) }}\ , \
		B_{44} = e^{\frac{K H_f}{(\lambda^{pre}) }}
	\end{gathered}
\end{equation}

\begin{figure}
	\centering
	\includegraphics[scale=0.9]{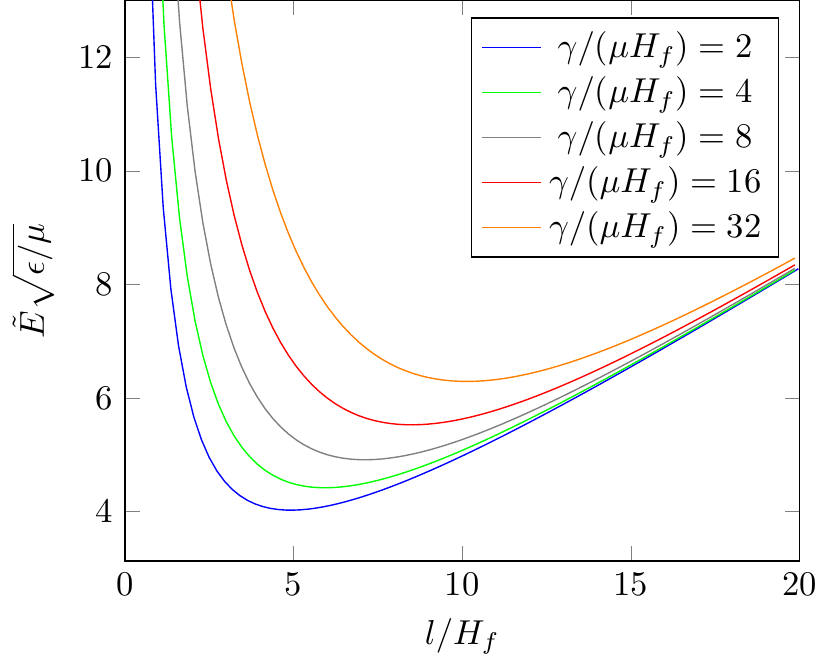}
	\caption{Analytic solution for $\lambda^{pre}=0.8$ (or $\varepsilon^{pre}=0.2$) shows normalized nominal electric field vs. normalized wavelength for various elastocapillary numbers $\gamma/(\mu H_f)$}
	\label{fig:e2_elec}
\end{figure}

The solution to the eigenvalue problem in (\ref{eq:eig_2}) gives the relation between the nominal electric field and the wavelength at a given stretch $\lambda^{pre}$ for various elastocapillary numbers. For instance, Figure \ref{fig:e2_elec} shows the calculated nominal electric-field $\tilde{E}\sqrt{\epsilon/\mu}$ for inducing the wrinkling instabilities in DEs for a uniaxial compression of $\varepsilon^{pre}=0.2$. The normalized nominal electric-field first decreases and then increases as the wavelength increases. The lowest electric-field in each curve gives the critical nominal electric-field $\tilde{E}_c\sqrt{\epsilon/\mu}$ for wrinkling instability.
\begin{figure}
	\centering
	\includegraphics[scale=0.9]{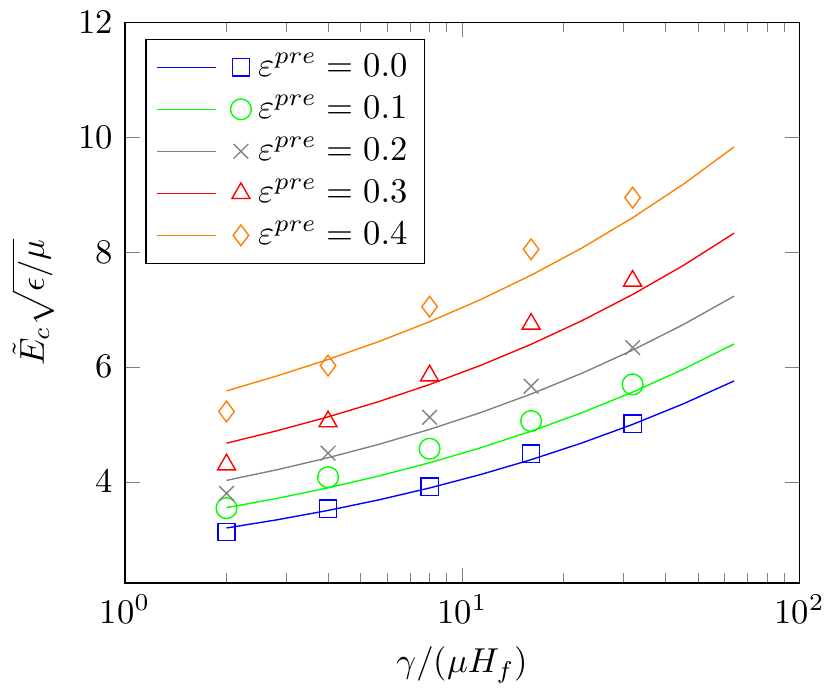}
	\caption{Critical electric field vs. elastocapillary number. The solid lines are the analytical results and the corresponding markers are the finite element solutions. }
	\label{fig:elecfig_plot}
\end{figure} 

We have recovered the analytic result in \citet{wang2013creasing} for wrinkling instability of a film without pre-compression $\varepsilon^{pre}=0$, as shown in Figure \ref{fig:elecfig_plot}.  The results in Figure \ref{fig:elecfig_plot} also demonstrate that pre-compressing the DE film leads to a significant increase in the electric field that is needed to induce the surface wrinkling instability, i.e. a near doubling of the voltage is needed as the initial compressive strain increases from 0 to 40\%. Pre-stretch has been widely observed to decrease the nominal electric field required for pull-in instability~\citep{zhaoAPL2007}, thus compression induces the opposite effect, that of increasing the nominal electric field $\tilde{E}$ required for electromechanical instability.

Increasing the surface tension, and thus the elastocapillary number has a similar effect, in that the critical nominal electric field needed to induce surface wrinkling increases substantially for a given pre-compression as the elastocapillary number increases. Again, a near-doubling of the critical electric field is needed as the elastocapillary number increases from 1 to 100.  

\begin{figure}
	\centering
	\includegraphics[scale=0.3]{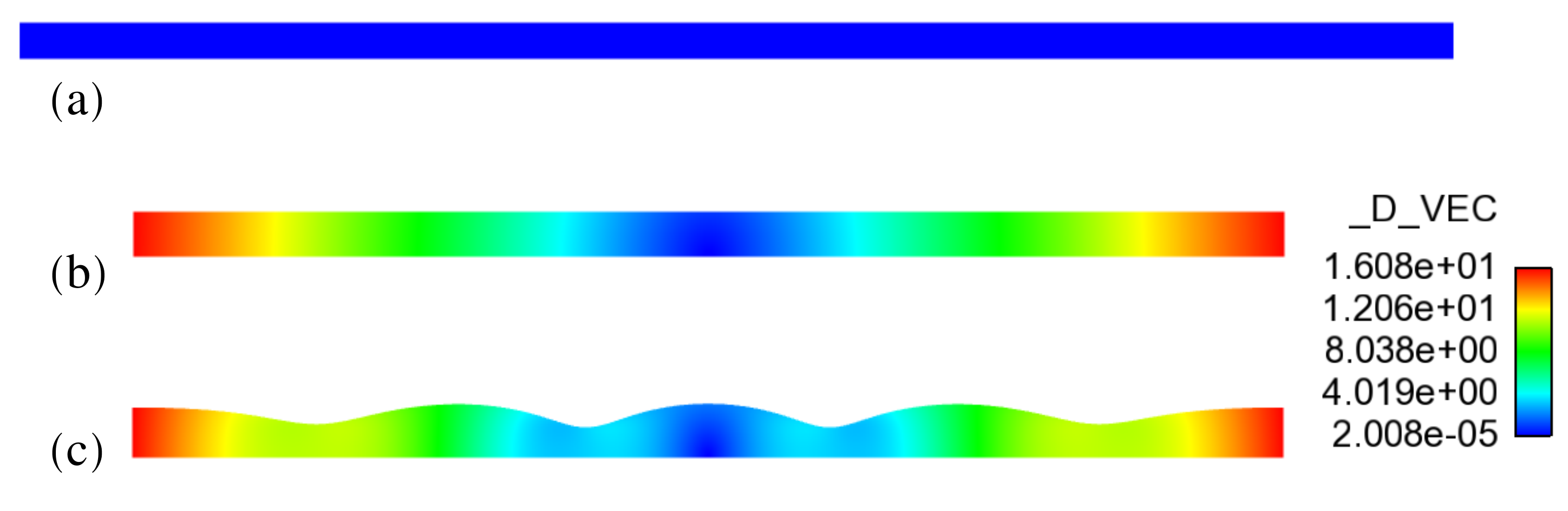}
\caption{FE simulation for electrically induced instability of a $160\times4$ DE film. (a) Undeformed film (b) Compressed film (in this case $\varepsilon^{pre}=0.2$) (c) Wrinkled structure due to applied electric field and surface tension (here $\gamma/(\mu H_f)=8$).}
	\label{fig:elec_fem}
\end{figure}

These results were verified using the nonlinear FE method of~\citet{seifiIJSS2016}.  To study the surface instability of a pre-compressed film subjected to electric field, we first statically compress a film with length of $L_f=160$ and height $H_f=4$ with mesh size $d=1/16$ to various strains of $\varepsilon^{pre}=0.1,\ 0.2,\ 0.3\ \text{and}\ 0.4$, all of which are smaller than Biot's wrinkling strain $\varepsilon_w=0.46$. Once the film is compressed, an electric potential on top of the film is applied (Figure \ref{fig:elecmechfig}-b) in conjunction with surface tension $\gamma$. The electric potential $\Phi$ was then increased linearly with time until the surface wrinkling instability occurs, where an illustration of the resulting surface wrinkling instability that is observed is shown in Figure \ref{fig:elec_fem}.  The normalized critical nominal electric field $\tilde{E}_c\sqrt{\epsilon/\mu}$ is then measured as soon as the wrinkles appears on the surface. In Figure  \ref{fig:elecfig_plot}, this critical electric field is plotted using symbols as a function of elastocapillary number $\gamma/(\mu H_f)$. The FE results are in a good agreement with our perturbation analysis. 

\begin{figure}
	\centering
	\includegraphics[scale=0.9]{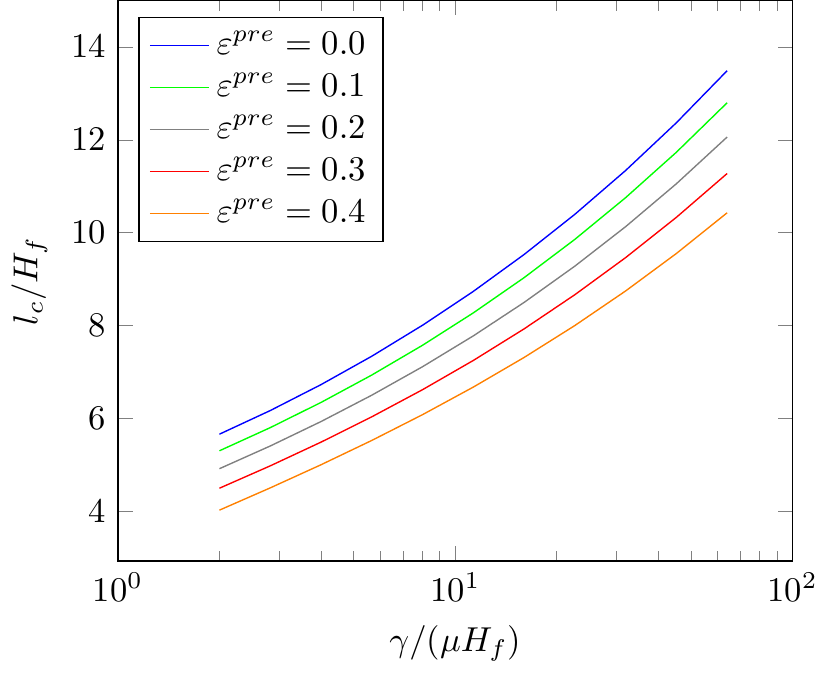}
	\caption{Analytic solution of critical wavelength vs. elasto-capillary number. }
	\label{fig:elecfig_plot2}
\end{figure}

Finally, Figure \ref{fig:elecfig_plot2} shows that for a given elastocapillary number, the wrinkling wavelength decreases with increasing DE film compressive strain.  This trend is opposite from Figure \ref{fig:mechfig_plot_fem}, where the wavelength increases with increasing elastocapillary number. This is because the pre-compressive strain $\varepsilon^{pre}$ reduces the surface area.  This reduction in surface area also reduces the surface energy, which results in a decrease in the critical wavelength.

\section{Conclusions}

In conclusion, we have presented a theoretical model augmented with computational analysis to examine the surface instabilities of dielectric elastomers accounting for surface tension effects. Surface tension is found to significantly impact the nature of the surface instabilities in the DEs, both through increasing the electric fields required to induce the instability, as well as in increasing the wavelength of the resulting instability.  We note that in the purely mechanical compression of soft materials, creases always occur before wrinkles regardless of the surface tension value~\citep{chenPRL2012}.  However, in electromechanically coupled soft materials like DEs, wrinkles occur before creases when the elastocapillary number is larger than one~\citep{wang2013creasing}.  Indeed, here we have focused on the analysis for elastocapillary numbers larger than one, as illustrated in Figs. \ref{fig:elec_fem} and \ref{fig:elecfig_plot2}.

\section{Acknowledgements}

HSP and SS acknowledge funding from the ARO, grant W911NF-14-1-0022.





\bibliography{biball,park_holmes_nsf_2016.bib}







\end{document}